\documentclass[preprint 2]{proto}
\usepackage{times}

\newcommand\pv{\mbox{$p_{V}$}}
\newcommand\pIR{\mbox{$p_{IR}$}}
\newcommand\irfactor{\mbox{$p_{IR}/p_{V}$}}

\begin{document}

\title{\textbf{\LARGE Space-Based Thermal Infrared Studies of Asteroids}}

\author {\textbf{\large Amy Mainzer}}
\affil{\textbf{\textit{\small Jet Propulsion Laboratory, California Institute of Technology}}}
\author {\textbf{\large Fumihiko Usui}}
\affil{\textbf{\textit{\small The University of Tokyo}}}
\author {\textbf{\large David E. Trilling}}
\affil{\textbf{\textit{\small Northern Arizona University}}}

\begin{abstract}
\begin{list}{ } {\rightmargin 1in}
\baselineskip = 11pt
\parindent=1pc
{\small Large-area surveys operating at mid-infrared wavelengths have proven to be a valuable means of discovering and characterizing minor planets.  Through the use of radiometric models, it is possible to derive physical properties such as diameters, albedos, and thermal inertia for large numbers of objects.  Modern detector array technology has resulted in a significant improvement in spatial resolution and sensitivity compared with previous generations of space-based infrared telescopes, giving rise to a commensurate increase in the number of objects that have been observed at these wavelengths.  Space-based infrared surveys of asteroids therefore offer an effective method of rapidly gathering information about small body populationsÕ orbital and physical properties.  The AKARI, WISE/NEOWISE, \emph{Spitzer}, and \emph{Herschel} missions have significantly increased the number of minor planets with well-determined diameters and albedos.
 \\~\\~\\~}

\end{list}
\end{abstract}

\section{\textbf{INTRODUCTION}}

A variety of remote sensing techniques have given rise to much of our understanding of the small body populations in our solar system.  Ground-based surveys operating at visible wavelengths have discovered most of the minor planets that are known today.  At present, roughly 670,000 asteroids have been discovered in the Main Belt between Mars and Jupiter, $\sim$12,000 near-Earth objects (NEOs) are known at all sizes, and some 5000 Jovian Trojans have been found.  While these numbers are thought to represent only a tiny fraction of the small bodies believed to exist in our solar system, much can be learned about these populations by studying the physical and dynamical properties of representative samples \citep[e.g.][and Jedicke et al. in this volume]{Bottke.2005a, Mainzer.2011e, Grav.2011a}.  Moreover, information about the probable composition and mineralogy of asteroids can be learned from visible and near-infrared (VNIR) spectroscopy and spectrophotometry.  Such studies have been carried out for $\sim$2500 asteroids \citep[e.g. the MIT-UH-IRTF Joint Campaign for NEO Spectral Reconnaisance,][and many others]{Tholen.1989a, Xu.1995a, Bus.2002a, DeMeo.2009a, Reddy.2010a, Kuroda.2014a}.  Compared to broadband imaging, spectrally resolved observations require brighter sources for the same signal-to-noise ratio, which limits the number of targets observable.  The fourth release of the Sloan Digital Sky Survey (SDSS) Moving Object Catalog \citep{Stoughton.2002a, Abazajian.2003a} provided $\sim$100,000 observations in $u$, $g$, $r$, $i$, and $z$ filters, leading to taxonomic classifications for $\sim$64,000 asteroids \citep{Carvano.2010a}.

Asteroid lightcurves, sometimes combined with stellar occultation data, have been inverted to obtain models of shapes and rotational states for hundreds of objects \citep[][see also Durech et al. in this volume]{Kaasalainen.2001a, Durech.2010a}.  Radar observations can produce the most detailed information about shape, size, orbit, and spin state short of visiting a body with spacecraft \citep[e.g.][and Benner et al. in this volume]{Ostro.2002a}.  Since radar echoes must be sent to and received from a body, sensitivity drops as distance to the fourth power, so objects must make fairly close approaches in order to be detected.  Approximately 600 asteroids have been observed with radar to date({\it http://echo.jpl.nasa.gov/asteroids/index.html}).  Polarimetry offers a means of studying asteroid surface properties, but since only a small fraction of the total luminosity is polarized, this technique has been applied to $\sim$300 asteroids thus far \citep{Lupishko.2012a, Gil-Hutton.2012a}.

While powerful, these techniques have only been employed on a small fraction of asteroids.  Until recently, for the vast majority of asteroids, nothing was known except for absolute visible magnitudes (denoted $H$) and orbital parameters.  Observing small bodies at thermal infrared (IR) wavelengths with space telescopes complements other techniques such as visible light ground-based surveys, VNIR spectroscopy and spectrophotometry, radar studies, and \emph{in situ} spacecraft visits.  If a telescope can be cryogenically cooled such that its sensitivity is limited by the natural backgrounds in space, rather than by self-emission, it is possible to obtain measurements of physical properties such as diameters and albedos very rapidly for a large number of asteroids.  Advances in IR detector technology have made it possible to achieve diffraction-limited imaging and orders-of-magnitude improvement in sensitivity compared to previous generations of IR space telescopes such as the \emph{Infrared Astronomical Satellite} \citep[IRAS;][]{Neugebauer.1984a, Tedesco.2002a}, the \emph{Mid-Course Space Experiment} \citep[MSX;][]{Mill.1994a, Price.2001a, Tedesco.2002b} and the \emph{Infrared Space Observatory} \citep[ISO;][]{Kessler.1996a}.  More recent surveys have now observed a substantial fraction of the known asteroids at thermal IR wavelengths, allowing for robust determinations of their diameters and (where corresponding visible light observations are available) albedos.  These measurements in turn inform the understanding of the dynamical and collisional history of the asteroids, their probable compositions and structure, and the impact hazard they pose to the Earth.  

Ground-based thermal IR observations began to play an important role in the physical characterization of the largest asteroids beginning in the 1970s with the advent of IR detectors \citep[e.g.][]{Allen.1970a, Allen.1971a, Matson.1971a}.  These observations require the largest ground-based telescopes due to the enormous thermal background from the Earth's atmosphere and the telescope itself (backgrounds $\sim10^{6}$ lower between $\sim$4-200 $\mu$m are achievable with a space-based cooled telescope).  Moreover, the Earth's atmospheric opacity restricts ground-based observations to selected wavelengths.  Some mid-IR instruments are available on ground-based telescopes, especially at high-altitude observatories in Hawaii and Chile, leading to observations of several hundred objects to date \citep[e.g.][]{Hansen.1976a, Cruikshank.1977a, Lebofsky.1978a, Morrison.1979a, Delbo.2003a, Delbo.2011a, Matter.2011a, Mueller.2012a, Mueller.2013a, Wolters.2005a, Wolters.2008a}.  

Thermal IR fluxes can be used to derive physical properties such as effective spherical diameter (defined as the diameter of a spherical thermal model asteroid emitting the same IR flux as that observed) through the use of radiometric models. Section 3 below and Delbo et al. (this volume) give a more detailed description of common thermal models; see also \citet{Harris.2002a}.  Small body spectral energy distributions consist of the sum of blackbody curves produced by reflected sunlight, centered at visible wavelengths, and blackbody curves generated by thermal emission from areas at different temperatures.  The central wavelength of the thermal peak depends on an object's heliocentric distance and thermal properties but usually lies somewhere between 5 -- 20 $\mu$m for most asteroids interior to Saturn's orbit.  Hence, by combining orbital information with thermal IR observations that bracket an object's rotational lightcurve, it is possible to constrain the total energy being emitted by an asteroid and convert this into an effective spherical diameter.  With only a single IR measurement, it is possible to compute a projected size, but this does not necessarily correspond to the effective spherical diameter if the object is elongated.  Diameters derived from good-quality thermal IR measurements are considerably more accurate than those estimated from visible light observations alone, since an asteroid's visible flux depends strongly on its albedo.  With multiple thermally-dominated IR bands that adequately sample rotational phase, effective spherical diameters can be determined to within $\pm$10\% \citep{Mainzer.2011b}; see Section 3 for a more detailed discussion.  However, since asteroid albedos range from a few percent to $\sim$50\% \citep[e.g.][]{Binzel.2004a, Mainzer.2011e}, effective spherical diameters derived from only visible light measurements are typically uncertain by factors of 2-3.  If both IR and visible observations are available, it is possible to solve for albedo as well.

Space-based thermal IR surveys therefore offer a number of advantages for understanding small body populations.  Since an asteroid's visible flux is a strong function of its albedo, visible light surveys are preferentially biased against low albedo objects, which are intrinsically dimmer than higher albedo asteroids.  By contrast, surveys that independently detect asteroids (whether previously known or new) based solely on their thermal IR flux are approximately equally sensitive to low and high albedo asteroids.  IR-selected samples directly measure diameter and are insensitive to albedo, allowing for direct determination of a population's size-frequency distribution rather than its $H$-magnitude distribution.  An IR-selected sample can therefore be readily extrapolated to determine the orbital and physical properties of the underlying population.  

In addition, albedos derived from the combination of visible and IR observations offer clues to an object's probable composition.  Accurate measurements of asteroid diameters are needed to constrain the impact energy of potentially hazardous objects and inform mitigation strategies.  Space-based telescopes can, depending on their design, observe regions of sky that are inaccessible to ground-based observers.  All-sky surveys offer the additional benefit of sampling the population of objects with high inclinations more thoroughly.  Thermal IR observations can be used to determine thermal inertia, a key parameter for understanding the nature of asteroid regoliths (see Delbo et al., this volume).  Depending on wavelength and spectral resolution, space-based IR instruments can allow for measurement of emission and absorption features that are difficult or impossible to access from ground-based facilities. 

\section{The Missions}
The IRAS space telescope was the first space mission to survey the minor planets at thermal IR wavelengths in 1983, followed by MSX and ISO in 1995--1998 and 1996--1997, respectively. With 62 pixels and a 60-cm telescope, the IRAS mission observed $\sim$2200 asteroids \citep{Matson.1989a, Veeder.1989a, Tedesco.2002a}.  ISO observed $\sim$40 different asteroids in targeted mode \citep{Dotto.2002a}, and MSX observed an additional $\sim$30 objects not detected by IRAS \citep{Tedesco.2002b}.  IRAS was not diffraction limited due to the large physical size of its pixels; its minimum spatial resolution was $\sim$30 arcsec.  IRAS discovered six asteroids, including (3200) Phaethon, the parent body of the Geminid meteor stream \citep{Whipple.1983a, Green.1985a}, and five comets.

As larger-format IR arrays with smaller pixels and lower noise became available, they were incorporated into next-generation observatories such as the \emph{Spitzer Space Telescope} \citep{Werner.2004a}, the AKARI mission \citep{Murakami.2007a}, the \emph{Wide-field Infrared Survey Explorer} \citep[WISE;][]{Wright.2010a}, and the \emph{Herschel Space Telescope} \citep{Pilbratt.2010a}.  These new arrays allowed for Nyquist-sampled images (i.e. the pixel width is less than half the size of the point spread function produced by the telescope), improving spatial resolution and sensitivity by orders of magnitude over IRAS and MSX.  

The advances in microelectronics, visible-light camera chips, and tactical sensors are directly responsible for the vast improvement in sensitivity and miniaturization from the IRAS mission to today.  The 62 IRAS detectors were individually manufactured and hand-assembled; each pixel was $\sim$1-2 mm across \citep{Young.1993a}.  Major developments since IRAS include the advent of monolithically fabricated arrays of pixels and the ability to produce semiconductor material with extremely low levels of impurities.  The result is the development of megapixel IR arrays capable of achieving low read noise and dark currents well below the natural zodiacal background at thermal IR wavelengths. For example, the WISE mission included four 1024$^{2}$ detector arrays, each with 18 $\mu$m pixels \citep{Mainzer.2005a}.  Recent developments include the fabrication of megapixel mid-IR detectors capable of operating at higher temperatures, which will ease the burden on cryogenic systems for future missions \citep{McMurtry.2013a}.

\emph{Spitzer}, AKARI, WISE, and \emph{Herschel} were cryogenically cooled during their prime mission phases, and all but \emph{Herschel} continued to operate using the wavelengths that remained available following depletion of their liquid or solid cryogens.  Table 1 summarizes the key technical and operational differences between them.  \emph{Spitzer} and \emph{Herschel} functioned as general-purpose observatories, with a wide range of observing modes employed during their fully cryogenic missions.  Both performed targeted observations of previously known objects; neither mission has been widely used for asteroid discovery.  WISE operated in a single survey mode, covering the entire sky in six months.  Modifications to its science data processing pipeline allowed new discoveries of minor planets to be made. The primary objective of the AKARI mission was to carry out an all-sky survey; AKARI also had the capability of performing targeted spectroscopic and imaging observations.  After the liquid cryogen boiled off, only near-IR observations in a targeted mode were carried out. AKARI did not discover new asteroids; the all-sky observing cadence did not support the multiple observations needed to discover new moving objects.

In this chapter, we review only space-based asteroid studies from remote telescopic observations.  For a detailed discussion of the results from mid-IR observations of small bodies carried out by \emph{in situ} missions such as Dawn and Rosetta, the reader is referred to chapters by Barucci et al. and Russel et al. (this volume).  These missions gathered thermal IR maps of bodies such as (2867) Steins \citep{Leyrat.2011a}, (21) Lutetia, and (4) Vesta \citep{Keihm.2012a, Tosi.2014a, Capria.2014a}.

\begin{deluxetable}{lllll}
\tabletypesize{\small}
\tablecaption{Summary of Recent Space-Based IR Telescope Capabilities \label{tbl-1}}
\tablewidth{0pt}
\tablehead{Mission &  Dates & Wavelengths & \begin{tabular}{l}Field of \\View (arcmin) \end{tabular}& \begin{tabular}{l}Observing \\Modes  \end{tabular}\\ }
\startdata
\emph{Spitzer} & 2003-2008 & \begin{tabular}{l}IRAC: Imaging 3.6, 4.5, 5.8, 8 $\mu$m \\IRS: Spectroscopy 5-35 $\mu$m\\IRS: Imaging 16, 22 $\mu$m\\MIPS: Imaging 24, 70, 160 $\mu$m\\MIPS: Spectroscopy 55 - 95 $\mu$m\end{tabular} & \begin{tabular}{l}IRAC: 5x5\\IRS: $\sim$0.9x1.35\\MIPS: 5x5 \end{tabular} & Targeted\\
\hline
\emph{Spitzer} & 2008 - present & \begin{tabular}{c}IRAC: Imaging 3.6 \& 4.5 $\mu$m \end{tabular}& \begin{tabular}{l}IRAC: 5x5\end{tabular} & Targeted\\
\hline
\hline
AKARI & 2006-2007 &  \begin{tabular}{l}IRC: Imaging 9, 18 $\mu$m \\FIS: Imaging 65, 90, 140, 160 $\mu$m \end{tabular} & \begin{tabular}{l}IRC: 10 (cross-scan direction) \\FIS: 8-12 \\(cross-scan direction) \end{tabular}& Survey\\
\hline
AKARI & 2006-2007 & \begin{tabular}{l}IRC: Imaging\\2.4, 3.2, 4.1, 7, 9, 11, 15, 18, 24 $\mu$m \\IRC: Spectroscopy 1.8-5.2, \\5.4-12.9, 17.5-25.7 $\mu$m\\FIS: Imaging 65, 90, 140, 160 $\mu$m \\FIS: Spectroscopy 60-180 $\mu$m \end{tabular}& \begin{tabular}{l} IRC imaging: 10 x 10 \\FIS imaging: \\8-12 (cross-scan direction) \end{tabular} & Targeted\\
\hline
AKARI & 2008-2011 & \begin{tabular}{l}IRC: Imaging 2.4, 3.2, 4.1 $\mu$m \\IRC: Spectroscopy 2.5-5 $\mu$m \end{tabular}& \begin{tabular}{l} IRC imaging: 10 x 10  \end{tabular} & Targeted\\
\hline
\hline
WISE & 2010-2011 & \begin{tabular}{l}Imaging 3.4, 4.6, 12, 22 $\mu$m \end{tabular} & \begin{tabular}{l}47x47\end{tabular} & Survey\\
\hline
NEOWISE & 2013-present & \begin{tabular}{l}Imaging 3.4 \& 4.6 $\mu$m \end{tabular} & \begin{tabular}{l}47x47\end{tabular} & Survey\\
\hline
\hline
\emph{Herschel} & 2009-2013 & \begin{tabular}{l}HIFI: Spectroscopy 157-212 \\ \& 240-625 $\mu$m \\PACS: Imaging at  70, 100, 160 $\mu$m\\PACS: Spectroscopy 55-210 $\mu$m\\SPIRE: Imaging 250, 350, 500 $\mu$m\\SPIRE: Spectroscopy 194-617 $\mu$m\\\end{tabular} & \begin{tabular}{l}PACS Imager: \\1.75x3.5\\PACS Spectrometer: \\47x47 \\SPIRE Imager: 4x8\\ Spectrometer: 2.6 \end{tabular} & Targeted \\ 
\enddata
\tablecomments{As of March 2015.}
\end{deluxetable}

\subsection{The Spitzer Space Telescope}
The \emph{Spitzer Space Telescope} was launched on August 25, 2003 into an Earth-trailing orbit \citep{Werner.2004a}.  As a NASA ``Great Observatory," it was designed to function as a facility observatory for the astrophysics and planetary communities.  The 85 cm telescope was launched warm, sitting atop a $\sim$350 L superfluid helium cryostat that enclosed the instruments and fine guidance sensor.  During its 5.5 year fully cryogenic mission, three instruments were available: the Infrared Array Camera \citep[IRAC; an imager operating at 3.6, 4.5, 5.8, and 8 $\mu$m;][]{Fazio.2004a}, the Multiband Imaging Photometer for \emph{Spitzer} \citep[MIPS; imaging and spectroscopy covering the wavelength range 24 - 160 $\mu$m;][]{Rieke.2004a}, and the Infrared Spectrograph \citep[IRS; low and medium resolution spectroscopy between 5 - 35 $\mu$m and imaging at 16 and 22 $\mu$m;][]{Houck.2004a}.  The IRAC instrument observed the 3.4 and 5.8 $\mu$m channels simultaneously, and the 4.5 and 8 $\mu$m channels simultaneously.  During the fully cryogenic portion of the \emph{Spitzer} mission, the telescope temperature was maintained below 8 K using vapor escaping from the superfluid helium cryostat; focal planes were maintained at 1.4 K to support operation of the longest wavelengths.  

In 2008, the liquid helium was depleted; at this point, the focal planes and telescope reached an equilibrium temperature near 29 K.  The 3.6 and 4.5 $\mu$m channels of IRAC continued to operate via passive cooling in an extended Warm Mission phase \citep{Storrie-Lombardi.2012a}.  Hundreds of asteroids throughout the Solar System were targeted using all three \emph{Spitzer} instruments during the fully cryogenic and Warm Mission phases \citep[e.g.][among others]{Emery.2006a, Trilling.2010a, Licandro.2012a}.  

Data from \emph{Spitzer} can be accessed through the \emph{Spitzer} Heritage Archive ({\it http://sha.ipac.caltech.edu}).  In general, proprietary periods are one year for Guest Observer programs.

\subsection{AKARI} Consisting of a 68.5 cm telescope, the AKARI satellite (meaning ``light" as opposed to representing an acronym) launched on February 21, 2006 into a Sun-synchronous polar orbit at 700 km altitude \citep{Murakami.2007a}. AKARI carried two instruments, the InfraRed Camera \citep[IRC;][]{Onaka.2007a} and the Far-Infrared Surveyor \citep[FIS;][]{Kawada.2007a}, covering the spectral ranges of 2--26 $\mu$m and 50--180 $\mu$m, respectively.  The whole telescope and instruments were cooled down to $\sim$6 K using 180 liters of superfluid helium and two sets of two-stage Stirling cycle mechanical coolers \citep{Nakagawa.2007a}.  AKARI's liquid helium supply lasted until August 26, 2007 and enabled 550 days of fully cryogenic operations (the AKARI ``cold mission").

During the cold mission phase, AKARI completed an all-sky survey at six bands: 9, 18, 65, 90, 140, and 160 $\mu$m \citep{Ishihara.2010a, Yamamura.2010a}.  The mid-IR part of the all-sky survey was conducted at two broad bands centered at 9 and 18 $\mu$m with the IRC.  More than 90\% of the sky was observed with both bands, and a large portion of the sky was observed more than three times.  The point source catalog of $\sim$877,000 objects was produced from the mid-IR images of the all-sky survey data \citep{Ishihara.2010a}.  An asteroid catalog was also constructed from the mid-IR survey data \citep{Usui.2011a}.  The 16-month cold mission allowed the inner edge of the Main Belt to be observed at least once \citep{Usui.2013a}.  AKARI detected point sources by identifying objects that were detected at least twice at the same fixed point on the sky, leaving transient sources to be identified with  the list of solar system bodies known at the time of data processing.  Most asteroids in the AKARI catalog were detected fewer than five times, with $\sim$25\% having one detection \citep{Usui.2011a}.  The two mid-IR bands observed different regions of the sky, separated by $\sim$25$^\prime$ in the cross-scan direction; thus, solar system objects were not observed simultaneously in both bands.  In total, 5120 known asteroids were identified with AKARI detections.  The list of derived sizes and albedos, known as the Asteroid Catalog Using AKARI (AcuA), is publicly available ({\it http://darts.isas.jaxa.jp/ir/akari/catalogue/AcuA.html}).  The IR fluxes of individual asteroids observed with AKARI will also be released in the near future.

The instruments on board AKARI performed both deep imaging and spectroscopy in the targeted observation mode, occasionally inserted into a continuous survey operation.  The targeted observations were planned in advance to maximize the all-sky survey coverage and efficiency.  A number of observations for a wide variety of astrophysical targets ranging from solar system objects to galaxies at cosmological distances were carried out in the targeted mode.  During the cold mission, AKARI made more than 5000 targeted observations, including a small serendipitous survey of asteroids \citep{Hasegawa.2013a}, thermal IR photometric observations of (25143) Itokawa \citep{Mueller.2014b} and (162173) 1999 JU$_{3}$ \citep{Hasegawa.2008a}, the targets of the JAXA Hyabusa and Hyabusa 2 sample return missions (see Yoshikawa et al. in this volume).

After the exhaustion of AKARI's liquid helium, the telescope and instruments remained at sufficiently low temperatures (below 50 K) owing to the mechanical cooler.  Targeted near-IR observations with the IRC were therefore able to be carried out in the AKARI ``warm mission" phase.  Low resolution spectroscopy was performed using the near-IR channel of the IRC from 2--5 $\mu$m.  The AKARI warm mission's science observations began in June 2008 and continued until May 2011.  The IRC carried out more than 12,000 targeted observations during the warm mission, including spectroscopy of 70 asteroids.  The AKARI data are available at ISAS/JAXA ({\it http://www.ir.isas.jaxa.jp/AKARI/Observation/}).

\subsection{WISE/NEOWISE} Launched on December 14, 2009 into a 525 km Sun-synchronous orbit, the WISE mission surveyed the entire sky at four infrared wavelengths (3.4, 4.6, 12 and 22 $\mu$m; denoted W1, W2, W3, and W4 respectively) using a 40 cm telescope \citep{Wright.2010a}.  The survey's scientific objectives were to find the cool stars and luminous galaxies using a single operational mode.  The survey strategy was designed for rapid sky coverage resulting in observations of the entire sky with an average depth of $\sim$10 exposures on the ecliptic plane (rising to thousands at the ecliptic poles) after six months.  All four wavelengths were imaged simultaneously using beamsplitters.  Augmentations to the WISE science data processing pipeline allowed for archiving of the individual single-frame exposures as well as mining of moving objects from the images in real time \citep{Mainzer.2011a}.  The fully cryogenic baseline mission was completed in July 2010, and the solid hydrogen cryogen used to cool the W3 and W4 detectors was partially exhausted on August 5, 2010, resulting in the loss of the W4 channel.  Operations continued using the W1, W2, and W3 arrays until the cryogen was fully exhausted on September 30, 2010.  The mission continued in a post-cryogenic phase with the goal of observing near-Earth objects (NEOs; asteroids and comets with perihelia $<$1.3 AU) until February 1, 2011 using the passively cooled W1 and W2 channels.  After this, it was placed into hibernation for 32 months.  During all phases of the prime mission, $>$158,000 asteroids were detected, including $\sim$34,000 new discoveries.    

The spacecraft was reawakened in August 2013 to begin an extended mission, known as NEOWISE (for near-Earth object + WISE), to discover and characterize the NEO population using the 3.4 and 4.6 $\mu$m channels, which remain operational via passive cooling to $\sim$75 K.  Survey operations resumed on December 21, 2013 and are expected to continue until early 2017.  To date, the spacecraft has observed $\sim$11,700 minor planets, including 270 NEOs, a rate of $\sim$0.7-0.8 NEOs per day \citep{Mainzer.2014b}.  

The WISE observational strategy typically resulted in $\sim$10-12 observations of most asteroids spaced over $\sim$36 hours.  However, the short length of the observational arc means that ground-based follow-up is necessary to secure orbits for new discoveries.  The worldwide community of professional and amateur observers has contributed to follow-up for NEOWISE NEO candidate discoveries, which are typically found with $R\sim$22 mag.  As an all-sky survey, NEOWISE candidates are frequently found at high declinations, regardless of  weather or lunar phase, posing unique challenges for follow-up observers.

All data from the WISE prime mission have been publicly released and are available through NASA's Infrared Science Archive (IRSA; {\it http://irsa.ipac.caltech.edu/Missions/wise.html}).  Derived physical properties for minor planets are being prepared for submission to NASA's Planetary Data System; until then, they are available through the individual NEOWISE team publications, i.e. \citet{Mainzer.2011d, Mainzer.2012c, Mainzer.2014a, Mainzer.2014b} for the NEOs; \citet{Masiero.2011a, Masiero.2012a, Masiero.2014a} for the MBAs; \citet{Grav.2011b, Grav.2012a, Grav.2012b} for the Hilda group and Jovian Trojans; and \citet{Bauer.2013a} for the scattered disk objects and Centaur populations.  Data from the NEOWISE reactivation will be publicly released annually through IRSA beginning in March 2015.

\subsection{Herschel} The \emph{Herschel Space Observatory} launched on May 14, 2009, carrying a 3.5 m telescope into orbit around the Earth-Sun L2 Lagrange point \citep{Pilbratt.2010a}.  \emph{Herschel's} telescope was passively cooled, and a dewar containing $\sim$2370 L of superfluid liquid helium was used to cool its far-IR instruments, the Photodetector Array Camera and Spectrometer \citep[PACS;][]{Poglitsch.2010a}, the Spectral and Photometric Imaging REceiver \citep[SPIRE;][]{Griffin.2010a}, and the Heterodyne Instrument for the Far Infrared \citep[HIFI;][]{deGraauw.2010a}.  Owing to its longer wavelengths, spanning 55 -- 671 $\mu$m, \emph{Herschel} was primarily used to observe the more distant minor planets (which are cooler and therefore emit strongly at far-IR) such as trans-Neptunian and Kuiper Belt objects by the open time key program entitled ``TNOs are Cool: A survey of the trans-Neptunian region" \citep{Mueller.2010a}.  However, Main Belt asteroids were used as calibration sources \citep{Mueller.2014a}, and measurements of individual targets of interest were made for $\sim$10 asteroids \citep[e.g.][]{Leyrat.2012a, ORourke.2012a, Mueller.2013a, Mueller.2014b, Mueller.2014c}.  

Herschel data may be accessed through the \emph{Herschel} Science Archive (HSA) maintained by the European Space Agency ({\it http://www.cosmos.esa.int/web/herschel/science-archive}).  It is also possible to query the HSA through NASA's \emph{Herschel} Science Center ({\it http://irsa.ipac.caltech. edu/applications/Herschel/}).

\begin{deluxetable}{llllllll}
\tabletypesize{\small}
\tablecaption{Summary of Recent Space-Based IR Asteroid Observations and Discoveries \label{tbl-2}}
\tablewidth{0pt}
\tablehead{Mission &  NEOs & MBAs & Cybeles & Hildas & Jovian Trojans & Centaurs &  \begin{tabular}{l}Asteroid \\Discoveries  \end{tabular} \\ }
\startdata
\emph{Spitzer} & $<$10 & $\sim$250 & 1 & 62 & $\sim$70 & 42 & 0 \\
\emph{Spitzer} Warm Mission & $\sim$600 & 0 & 0 & 0 & 0 & 28 & 0\\
AKARI & 90 & 4806 & 106 & 86 & 110 &7 & 0 \\
WISE & $\sim$700 & $\sim$158,000 & 1342 & 1023 & $\sim$2000 & 52 & $\sim$34,000\\
NEOWISE & 270 & $\sim$11600 & 100 & 50 & 50 & 1 & $\sim$100\\
\emph{Herschel} & $\sim$7 & $\sim$8 & 1 & 0 & 0 & 18 & 0 \\
\enddata
\tablecomments{As of March 2015.}
\end{deluxetable}

\section{Thermal Modeling}
Thermal models range in sophistication and complexity; different models are used depending on data quality, computing resources, and the availability of ancillary data such as shape models and rotational states.  All models begin by solving the energy balance between incident solar radiation, reflected sunlight, and thermal emission \citep[e.g.][]{Lebofsky.1986a}. For a spherical, airless body, this is given by \begin{eqnarray}  A_{p} S = L_{r} + L_{e}, \end{eqnarray} where $A_{p}$ is the objects's projected area, $S$ is the solar flux at the distance of the asteroid, and $L_{r}$ and $L_{e}$ are the reflected and emitted radiation, respectively, and $L_{r}$/$L_{e}$ = $A/(1-A)$, where $A$ is the bolometric Bond albedo.  In essence, the purpose of thermal modeling is to determine $L_{e}$ from the observed flux in one direction (or a few directions for objects observed at multiple viewing geometries), and one or a few thermally-dominated bandpasses. The extrapolation to all directions and the bolometric emitted radiation are what the model provides. The total thermal emission is given by \begin{eqnarray}L_{e} = \epsilon \eta \sigma R^{2} \int^{\pi}_{-\pi} \int^{\pi/2}_{-\pi/2} T^{4}(\theta, \phi) cos(\phi) d\phi d\theta, \end{eqnarray} 

 where $\epsilon$ is the emissivity as a function of wavelength, $\eta$ is the so-called beaming parameter (described in more detail below), and $\sigma$ is the Stefan-Boltzmann constant, $R$ is the object's radius, and $T$ is the object's temperature distribution as a function of longitude ($\theta$) and latitude ($\phi$), measured from the subsolar point.  %

The Standard Thermal Model \citep[][and references therein]{Lebofsky.1986a} and the popular Near-Earth Asteroid Thermal Model \citep[NEATM;][]{Harris.1998a} adopt the following temperature distribution across the asteroid's surface: 

\begin{eqnarray}T(\theta, \phi) =  \Bigg\{ \begin{tabular} {ll} $T_{ss} cos^{1/4}\zeta$ & if $\zeta < \pi/2$  \\ 0 & if $\zeta \geq \pi/2$ \end{tabular} \end{eqnarray} where the angular distance from the subsolar point is $\zeta$ and the temperature at the subsolar point ($T_{ss}$) is \begin{eqnarray} T_{ss} = \left[ \frac{S(1 - A)}{\eta \epsilon \sigma} \right]^{1/4}. \end{eqnarray}

Since the peak of the Sun's spectral energy distribution occurs at visible wavelengths, the Bond albedo is customarily assumed to be equal to the total Bond albedo at $V$ band ($\sim$0.56 $\mu$m), $A_{v}$.  In the system of \citet{Bowell.1989a}, the Bond albedo is related to the more readily measured visible geometric albedo $p_{V}$ by  \begin{eqnarray}A \sim A_{v} = p_{V} q = p_{V}(0.29 + 0.684 G),\end{eqnarray} where $G$ is the so-called phase slope parameter that ranges from $\sim$-0.1 to 0.4 \citep{Bowell.1989a, Harris.1989a}.  

Since $p_{V}$ is generally $<$1, $A$ is usually $\ll1$.  We therefore have a direct relationship between the asteroid's thermal flux and its diameter $D$: $L_{e}\propto D^{2} (1-A) \rightarrow D^{2}$.  However, since the asteroid's reflected sunlight goes as $L_{r} = A_{p} A S \propto A D^{2}$, diameter and albedo cannot be easily disentangled if only $L_{r}$ is measured. This results in a much larger uncertainty in the derived diameter unless \pv is already known, given the large range of asteroid albedos \citep[$\sim$0.02 to $>$0.5;][]{Binzel.2004a, Mainzer.2011d, Masiero.2011a}.  

If only the absolute visible magnitude $H$ of the asteroid is available, corresponding to its V-band magnitude measured or extrapolated at $\alpha=0^{\circ}$ and object-to-observer and heliocentric distances of 1 au, a frequently-used empirical relationship for diameter is given by \begin{eqnarray} D = \frac{1329}{\sqrt{p_{V}}}10^{-H/5},    \end{eqnarray} where $D$ is the diameter in km  \citep{Bowell.1989a, Harris.2002a}.  This relationship also demonstrates that diameter derived from visible light observations alone is a sensitive function of an asteroid's albedo, underscoring the value of obtaining diameters obtained radiometrically.    

Thermal models offer the possibility of deriving physical properties for large numbers of asteroids, with increasingly sophisticated models and better data allowing for more and improved constraints on parameters such as diameter, geometric visible albedo (\pv), infrared albedo (\pIR), emissivity, shape, rotational state, and thermal inertia.  The STM dcorresponds to the case of a non-rotating spherical body, or one with zero thermal inertia and no night-side emission, observed at 0$^{\circ}$ solar phase angle.  The beaming parameter $\eta$, which takes into account surface roughness and the ``beaming" of thermal emission in the direction of the Sun, is often set to 0.756 in the STM based on empirical fits to the diameters of (1) Ceres and (2) Pallas derived from stellar occultations.  The fast rotating model \citep[FRM;][]{Lebofsky.1978a, Veeder.1989a, Lebofsky.1989a}, by contrast, is more appropriate for an object that rotates rapidly or has high thermal inertia/high thermal conductivity; effectively, $\eta$ is set to $\pi$, and the nightside emits flux and does not follow Equation 3.  The NEATM assumes a hybrid approach that still assumes zero contribution from the nightside, but allows $\eta$ to be fit as a free parameter if observations from more than one thermal IR band are available.  

NEATM-derived diameters generally reproduce measurements from radar, stellar occultations, and \emph{in situ} spacecraft visits to within $\pm$10\%, given multiple thermally-dominated IR measurements that adequately sample an asteroid's rotational light curve with good signal-to-noise ratio (SNR) and an accurate determination of distance from knowledge of its orbit \citep{Mainzer.2011b}.  It is worth noting that the accuracy of the diameters of objects used to confirm the performance of radiometric thermal models (such as radar or stellar occultations) is typically $\sim$10\%.  With these caveats, albedos can be determined to within $\pm$25\% of their value (i.e. \pv=0.04$\pm$0.01) if good-quality visible light observations are available.  The accuracy of the derived albedo depends critically on the accuracy of $H$; see below. If only observations at a single band centered near 4.6 $\mu$m are available (as is the case for some WISE/NEOWISE and \emph{Spitzer} targets), diameter errors typically degrade to $\pm20-25\%$, and albedos can only be known to within $\pm40-50\%$ of their value \citep{Mueller.2007a, Harris.2011a, Mainzer.2012c, Masiero.2012a}.    

While diameter determinations from thermal IR observations are relatively insensitive to visible light measurements, the latter are required to determine \pv. Obtaining good-quality $H$ and $G$ measurements \citep{Bowell.1989a} is a persistent difficulty.  Most of the time, visible observations collected simultaneously with IR fluxes are not available, so the absolute magnitude $H$ and phase curve slope parameter $G$ must be used to extrapolate the apparent visible magnitude at the time the IR observations were taken.  Most asteroid observations come from the visible-light surveys that discover them, such as the Catalina Sky Survey, LINEAR, and PanSTARRS.  Typically, discovery observations are made with broadband $V+R$ band filters to maximize sensitivity to NEOs, and while astrometric calibrations can be very accurate, photometric measurements can be considerably more variable.  Moreover, $G$ is known to vary with asteroid taxonomic type \citep[e.g.][]{Harris.1989b}.  \citet{Williams.2012a} recomputed $H$ and $G$ values for the entire Minor Planet Center catalog on a survey-by-survey basis and found that $H$ offsets peak at $+0.4$ mag in the range $H$=14.2 -- 14.5 mag, but decrease to $+0.1$ mag for $H>20$ mag, significantly expanding on the work of \citet{Pravec.2012a}, who measured $H$ values for 583 minor planets.  \citet{Williams.2012a} recomputed photometric magnitudes to account for the various surveys' filters.   NEOWISE thermal model fits were performed assuming large ($\pm$0.3 mag or more) errors for $H$ if no direct measurements of these parameters were available \citep[e.g. from][]{Warner.2009a, Pravec.2012a} at the time of publication. Future work will incorporate the \citet{Williams.2012a} $H$ and $G$ values into thermal models from NEOWISE.  If $H$ values are well-known, albedo can be determined to within 25\%, but if $H$ is poorly known, then the accuracy of the albedo degrades accordingly.

In bands where reflected sunlight contributes a non-negligible fraction of the total flux, the albedo at that wavelength (\pIR) must also be treated as a free parameter in thermal models.  For NEOs with subsolar temperatures close to $\sim$300 K, peak thermal emission occurs near $\sim$10 $\mu$m.  Wavelengths longer than $\sim$4 $\mu$m are typically thermally dominated for asteroids at heliocentric distances less than $\sim$4 AU; shorter wavelengths are a mix of reflected sunlight and thermal emission.  

The infrared albedo at $\sim$3 $\mu$m is not necessarily equal to the visible albedo, and it is correlated with taxonomic type.  The trends for each taxonomic type are \citep{Mainzer.2011d, Mainzer.2012a}: \begin{itemize} \item \irfactor $\sim$1 for C-complex asteroids (except for D-types)  \item \irfactor $\sim$1.7 for S-complex asteroids  \item \irfactor $\sim$2.2 for D-types \item \irfactor $\lesssim$1 for B-types \end{itemize} The differences between \irfactor\ for all of these taxonomic types is likely due to the fact that their VNIR slopes are either blue (as is the case for the B-types), flat (C-complex), red (S-complex), or very red (D-types), and the trend of these slopes continues out to $\sim$3$\mu$m.  Consequently, if \pIR\ can be determined, it can be used as a proxy for distinguishing between taxonomic types that otherwise have nearly identical visible albedos \citep{Grav.2012a, Grav.2012b, Ali-Lagoa.2013a, Masiero.2014a}.  

While the STM, FRM, and NEATM are useful tools for rapidly determining effective spherical diameter, \pv, and \pIR\ for large numbers of asteroids, they are of limited use for determining additional parameters such as emissivity and thermal inertia.  To extract these parameters, \emph{thermophysical} models are needed; see Delb{\'o} et al. (this volume) for a detailed discussion of their theory and application. 

\section{Results}

Table 2 summarizes the numbers of asteroids that are known to have been observed to date by \emph{Spitzer}, WISE/NEOWISE, AKARI and \emph{Herschel} for NEOs, Main Belt asteroids, Jovian Trojans, and Centaurs.  Given the capabilities of each mission's survey data and selection methods, much can be learned about the properties of the various asteroid populations.

\subsection{IR-Selected and Optically-Selected Population Studies.}
Samples of asteroids imaged by space-based IR telescopes can be selected in two ways.  In an independent survey, all small body candidates are treated identically regardless of whether they are previously known or might be new discoveries (assuming the survey's cadence allows for an observational arc sufficient to enable the discovery of new moving objects).  In a targeted sample, the telescope observes previously known asteroids and cannot independently discover new ones.

In an independent survey, asteroids are selected based on their IR fluxes, whereas in a targeted survey, asteroids are drawn from the catalog of objects discovered by other observers, almost all of whom operate at visible wavelengths.  Understanding the selection biases of the sample is critical when probing population properties below the size regime for which the sample is observationally complete.   

Since IR flux is insensitive to albedo, IR surveys are less biased against low albedo objects than surveys that select targets based on their visible flux alone.  Visible light surveys are less likely to discover low albedo objects, particularly at smaller sizes, due to their intrinsic faintness.  Moreover, since diameter is determined directly from IR fluxes, no conversion between $H$ and diameter is needed, eliminating the uncertainty associated with the large range of possible asteroid albedos.  IR-selected samples can be extrapolated to determine the size-frequency distribution of the underlying population they represent, since the sample selection method is insensitive to albedo.  

The NEOWISE project operated as an independent survey, using data processing methods to select moving object objects based on their IR fluxes.  Known and new candidates were treated identically by the WISE Moving Object Processing System (WMOPS), which required all moving object candidates to be detected a minimum of five times over $\sim$15 hours \citep{Mainzer.2011a, Cutri.2012a}. A detailed description of WMOPS can be found in the WISE Explanatory Supplement, section IV.5.  Most objects were observed an average of $\sim$12 times over $\sim$36 hours.  Over the course of the year-long prime mission, $\sim$158,000 asteroids were detected, including $\sim$34,000 new discoveries.  During the 8.5-month fully cryogenic portion of the mission, asteroids were selected by WMOPS based on their 12 $\mu$m flux; after the cryogen depleted, the primary band for asteroid selection became the 4.6 $\mu$m channel.

\begin{figure*}
\figurenum{1}
\epsscale{1.5}
\includegraphics[width=6in]{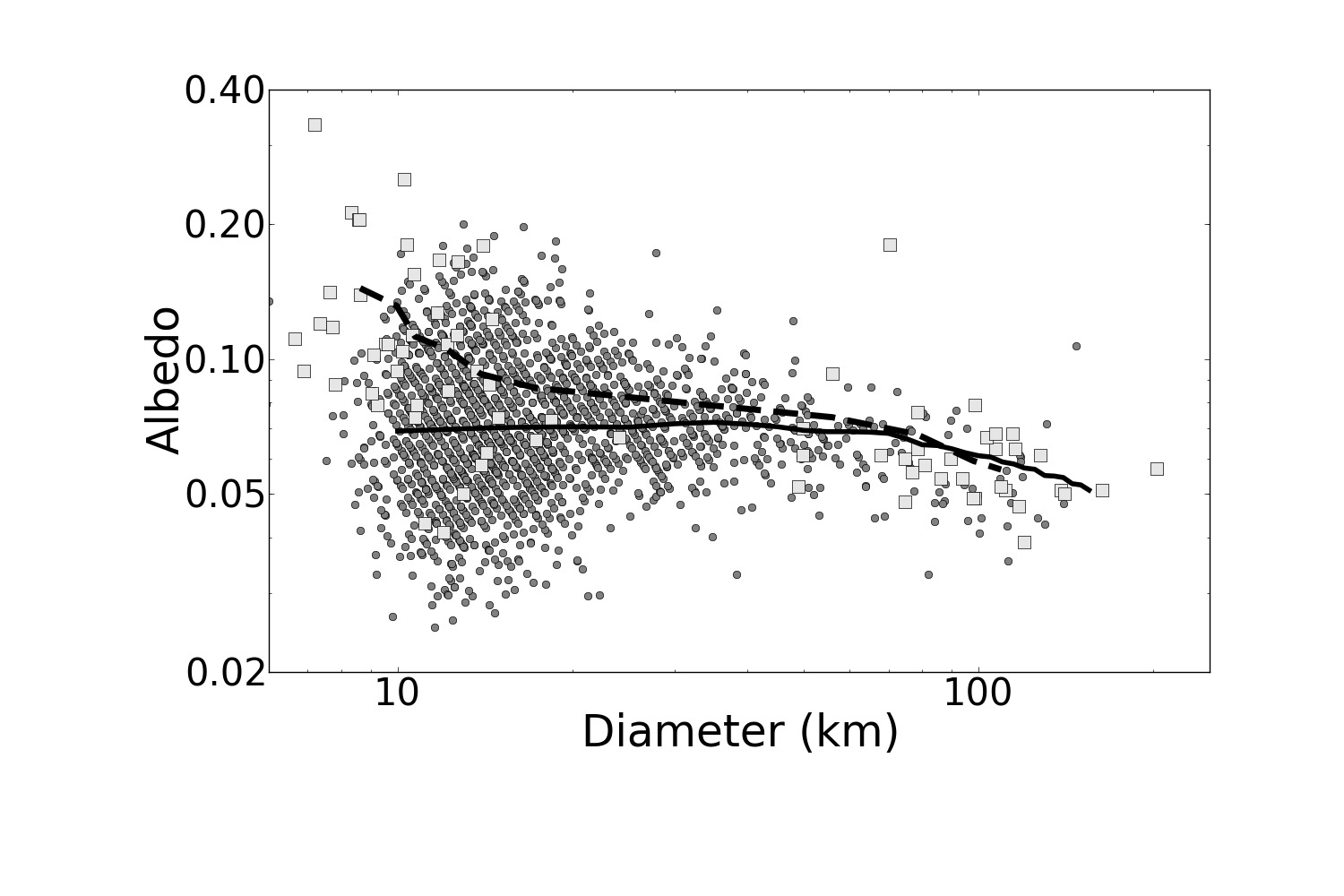}
 \caption{\small \label{fig:Trojans} Comparison of the IR-selected NEOWISE sample of 1739 Jovian Trojans  \citep[filled gray circles;][]{Grav.2011a, Grav.2012b} to the optically-selected sample observed by \emph{Spitzer} \citep[light gray squares][]{Fernandez.2009a}; the solid and dashed lines give the running medians for the NEOWISE and \emph{Spitzer} samples, respectively. }   
 \end{figure*}

\begin{figure*}
\figurenum{2}
\includegraphics[width=3in]{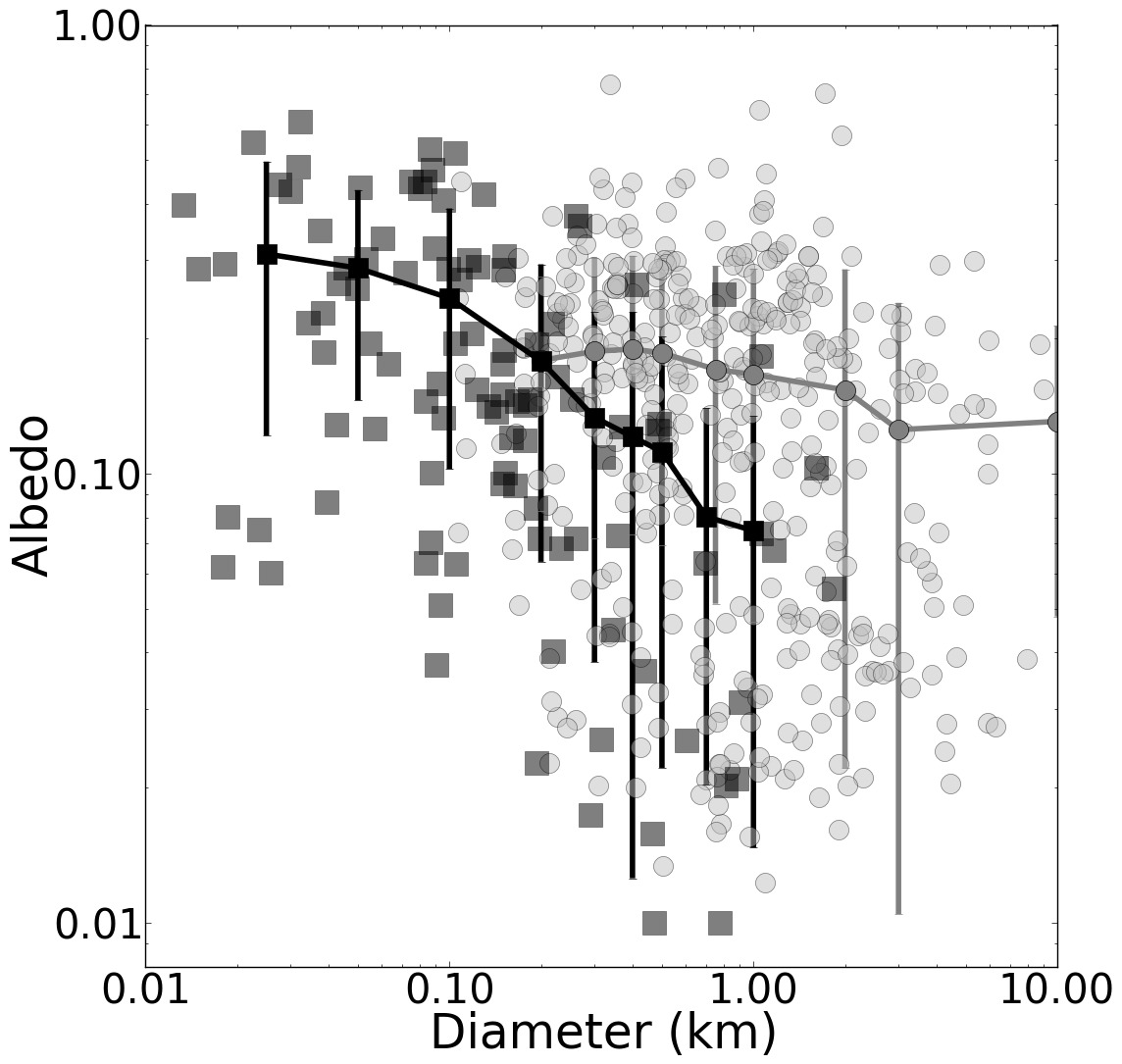}
\caption{\label{fig:diam_alb} The IR-selected sample of NEOs from NEOWISE (gray points; running median shown as gray line) shows little change in albedo with diameter, whereas a sample of asteroids selected based on their visible magnitudes (black points and line) shows an increase in albedo with decreasing diameter \citep{Mainzer.2014a}. }  
\end{figure*}

\begin{figure*}
\figurenum{3}
\includegraphics[width=3in]{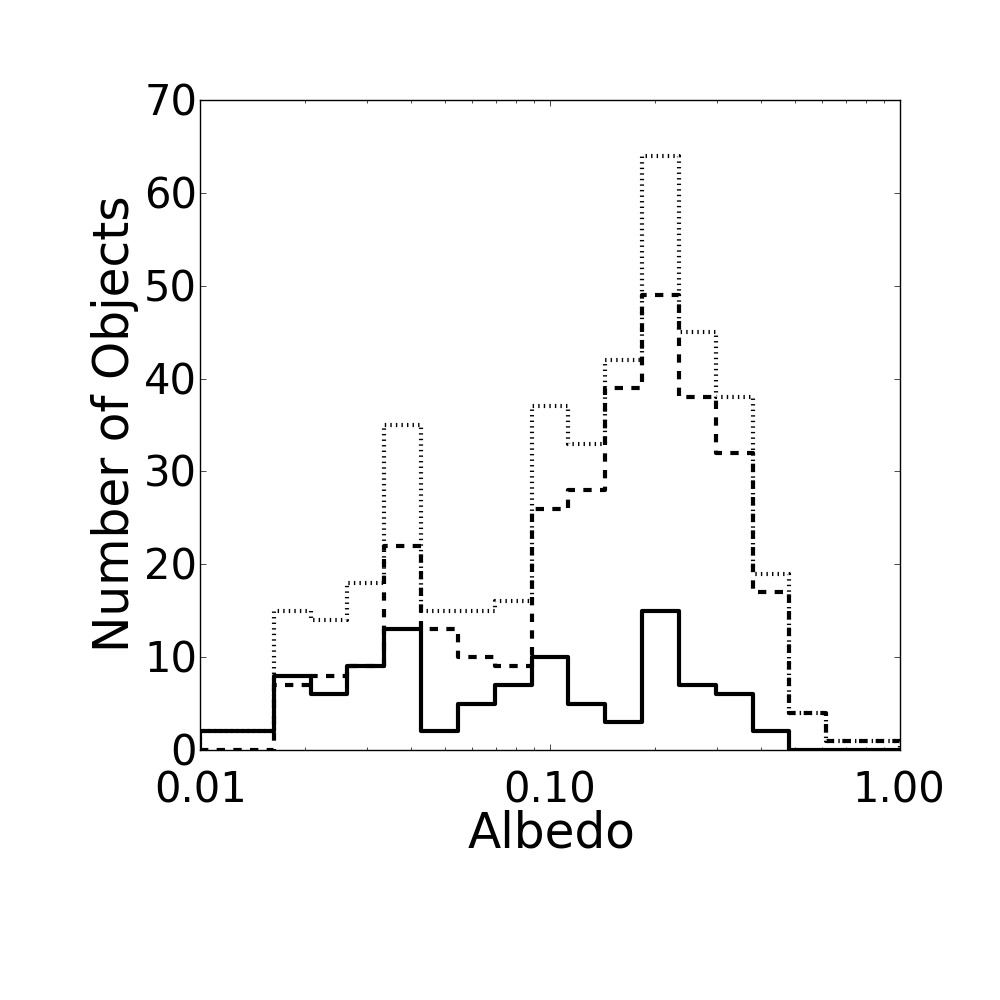}
\caption{\label{fig:NEOalbedos} The albedo distribution of the NEOs discovered by WISE during its prime mission (black) has relatively more low-albedo NEOs than the objects discovered by visible light surveys (dot-dashed line). The albedo distribution for the entire NEO sample from the WISE prime mission is shown as a finely dashed line.} 
\end{figure*}

One of the main results of the WMOPS-selected sample was the discovery that asteroid albedo distributions appear to remain constant over a wide range of diameters.  An example of this can be seen in Figure \ref{fig:Trojans}, which compares the albedo-size relationship for a sample of 44 optically-selected Jovian Trojans observed with \emph{Spitzer} \citep{Fernandez.2009a} to a sample of 1739 Trojans found using WMOPS \citep{Grav.2011a}.  While the optically-selected sample shows a strong trend of increasing albedo with decreasing size, no strong trend is apparent in the IR-selected, albedo-insensitive sample.  A similar effect is observed when the diameter-albedo relationship for NEO samples is examined \citep[Figure \ref{fig:diam_alb};][]{Delbo.2003a, Binzel.2004a, Trilling.2010a}.  For the small NEOs, a real increase in albedo with decreasing diameter cannot be ruled out, but because the sample was selected on the basis of its visible flux, any real trend is entangled with the optical surveys' bias against discovery of small, dark NEOs \citep{Mainzer.2014a}.  The NEOWISE-discovered NEOs have lower albedos than those discovered by optical surveys (Figure \ref{fig:NEOalbedos}). 

Using the IR-selected sample from NEOWISE, survey biases were estimated in order to determine the orbital properties, numbers, and physical properties of various asteroid populations.  Survey biases were computed as a function of orbital and physical properties using the complete survey pointing list and the sensitivity in each wavelength, which derives from observations of asteroids with well-determined orbits; see \citet{Mainzer.2011d} and \citet{Grav.2011a} for an expanded description of the debiasing methodology as applied to WISE data.  The process of removing the survey bias signatures from the observed sample is known as debiasing \citep[c.f.][]{Bottke.2002a, Jedicke.1998a, Spahr.1998a}.  

\citet{Mainzer.2011d} found that there are $\sim$20,500$\pm$3000 near-Earth asteroids larger than 100 m in diameter, of which $\sim$25\% had been discovered as of 2011.  That work verified that $>$90\% of NEAs larger than 1 km had been discovered, fulfilling the 1998 ``Spaceguard" goal given to NASA.  The sub-populations of the NEAs, including potentially hazardous asteroids (PHAs; asteroids whose minimum orbit intersection distances are $<$0.05 AU and $H$$<$22 mag), were studied by \citet{Mainzer.2012b}.  However, since WISE measures diameter, rather than $H$, the authors recommended a change to a diameter-based definition for PHAs and concluded that there are $\sim$4700$\pm$1500 PHAs larger than $\sim$100 m \citep{Mainzer.2012b}.  Matches to the orbital element models of \citet{Bottke.2002a} and \citet{Greenstreet.2013a} are generally in good agreement.  Approximately twice as many PHAs were found occupying orbits with the lowest inclinations as compared with the model of \citet{Bottke.2002a}; however, by recomputing the evolutionary models of NEOs with a finer orbital element grid, \citet{Greenstreet.2013a} found similar results for the lowest inclination bins.  NEOWISE did not effectively probe the NEA population below 100 m, with the exception of the optically-selected sample shown in Figure \ref{fig:NEOalbedos}.  Debiasing for this sample requires accounting for the optical survey selection effects in addition to the NEOWISE sample biases, a more complicated undertaking.    

\citet{Grav.2011b} and \citet{Grav.2012a} debiased the NEOWISE Jovian Trojan and Hilda samples, respectively, facilitating comparison between the leading and trailing Trojan clouds and leading to estimates of their numbers, size, and albedo distributions.  Infrared observations have revealed key attributes of the Jovian Trojans, Hildas, and Cybele asteroids.  These results indicate that Trojans in the leading and trailing clouds are extremely similar in terms of albedos, taxonomic types, and size distributions; the major difference is that Trojans in the leading (L4) cloud outnumber the trailing (L5) by a factor of 1.4$\pm$0.2 \citep{Grav.2011a}.  Theories of early giant planet migration must account for these observational constraints.  \citet{Marsset.2014a} obtained visible and near-IR spectra of eight Jovian Trojans found by WISE and \emph{Spitzer} to have relatively high albedos and found that these objects had taxonomic types consistent with other primitive, low albedo Trojans.  One likely explanation is that when determining albedos for thousands of objects, normal statistical fluctuations will scatter a small fraction of the observations in favor of brighter albedos.  See Emery et al. (this volume) for a more comprehensive discussion of the Jovian Trojans.

The \emph{Spitzer Space Telescope} observed an optically-selected sample of $\sim$600 NEOs during its Warm Mission at 3.6 and 4.5 $\mu$m \citep{Trilling.2010a, Mueller.2011a}.  Improved diameter constraints assist with assessments of impact hazard.  Targets of interest, such as (101955) Bennu, the target of the upcoming OSIRIS-REx mission, have also been observed with \emph{Spitzer} and \emph{Herschel} to determine size, albedo, and thermal inertia \citep[][and Delb{\'o} et al., this volume]{Emery.2014a, Yu.2014a, Mommert.2014a, Mommert.2014b}.  The target of the Hayabusa-2 mission, (162173) 1999 JU$_{3}$, has been intensively observed with AKARI, \emph{Spitzer}, and \emph{Herschel}, as well as ground-based facilities such as Subaru \citep{Hasegawa.2008a, Mueller.2011a}. WISE, \emph{Spitzer}, and \emph{Herschel} have been used to study Centaurs, a transitional population between trans-Neptunian objects and Jupiter-family comets \citep{Bauer.2013a, Duffard.2014a}.  Measurements from IRAS, AKARI, and WISE have been used to identify NEOs likely to be of cometary origin by performing follow-up observations of objects with low albedos and comet-like orbits \citep{Kim.2014a}. This work builds on the work of \citet{Demeo.2008a} and \citet{Fernandez.2005a}, who estimated the fraction of NEOs likely to be of cometary origins as 8\% and 4\%, respectively.  

The albedo-insensitive sample from \citet{Mainzer.2012a} suggests a paucity of NEOs with low albedos at low perihelion distances, although the sample size is small.  \citet{Delbo.2014a} posit that thermal cracking plays a key role in regolith production, with an erosion and destruction process that preferentially affects more fragile, low-albedo, carbonaceous objects as they approach the Sun; see also \citet{Capek.2010a, Capek.2012a}.

\subsection{Main Belt Asteroid Studies}
\begin{figure*}
\figurenum{4}
\includegraphics[width=3in]{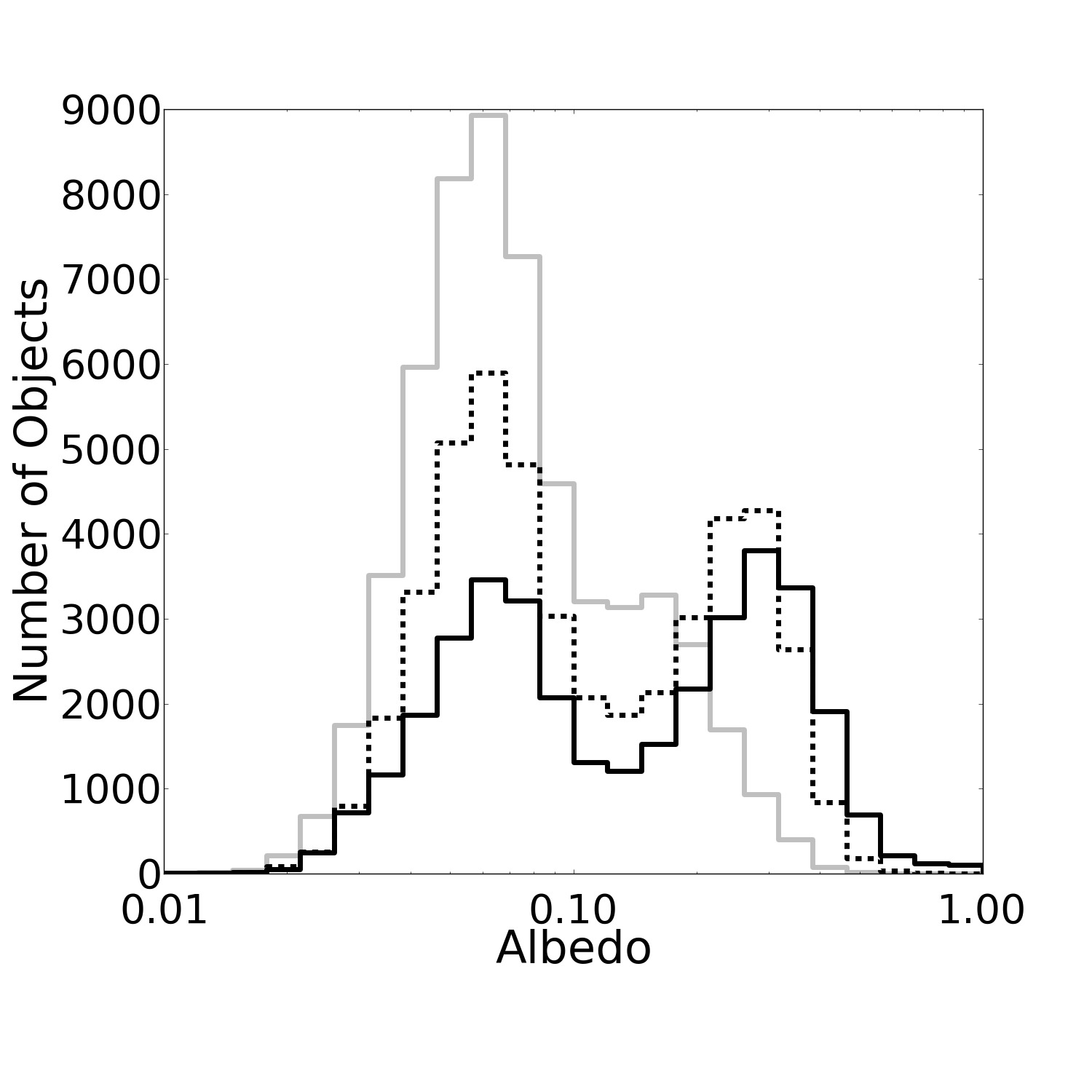}
\caption{\label{fig:MBAalbedos} The albedo distribution for asteroids in the inner (solid line; defined as semi-major axes between 1.8 and 2.5 au), middle (dashed line; 2.5 - 2.82 au), and outer (dotted line; 2.82 $<$ 3.6 au) regions of the Main Belt from \citet{Masiero.2011a}. } 
\end{figure*}

The Main Belt asteroids (MBAs) have now been extensively studied at thermal IR wavelengths, with $>$155,000 having been observed by WISE and 4806 objects by AKARI \citep{Masiero.2011a, Masiero.2012a, Usui.2011a, Hasegawa.2013a}.  \citet{Usui.2014a} shows the comparison of diameters and albedos derived for the $\sim$1900 MBAs observed in common between AKARI and WISE.  Diameters between the two datasets match to within $\pm$10\%, and albedos to within $\pm$22\%; these results are in good agreement with the error bars described above.  Figure \ref{fig:MBAalbedos} shows that, as found by IRAS, the asteroids within the Main Belt become darker with increasing heliocentric distance \citep{Masiero.2011a}.  

Radiometrically-derived diameters and albedos are valuable for improving determinations of asteroid collisional family ages and membership; in most cases, calculations prior to the recent IR missions described here have relied on sizes estimated from $H$ and an assumed albedo \citep[c.f.][]{Marzari.1995a}.  By combining albedos and colors with orbital information, it is possible to identify probable family members at velocity limits that would otherwise be indistinguishable from background objects, assuming that collisional family members have similar albedos and colors.  Furthermore, families that overlap in velocity space can be disentangled on the basis of their albedos and colors \citep{Masiero.2013a, Walsh.2013a, Carruba.2013a, Milani.2014a}.  With improved and expanded family membership lists down to smaller sizes, the timing of collisions can be determined with greater precision, since the age-determination technique relies upon the strength of non-gravitational forces such as the Yarkovsky and YORP effects, which in turn depends on objects' size and mass \citep{Masiero.2012b}.  Asteroid collisional families are discussed in greater depth in Nesvorny et al., Michel et al., and Masiero et al. in this volume.  

Diameters and albedos play a role in understanding which source reservoirs among the Main Belt asteroids and comets are most likely to have produced NEOs.  While it is difficult to associate the origins of an individual asteroid to a specific source region, associations can sometimes be inferred by comparing source region albedos to the object of interest and by using diameters to constrain migration timescales due to non-gravitational forces \citep[e.g.][]{Campins.2013a}.  Gathering measurements of albedos and diameters for larger numbers of asteroids down to smaller sizes will facilitate improved understanding of NEO origins and subsequent evolution, since non-gravitational forces depend strongly on asteroid sizes (see Vokrouhlicky et al. in this volume).

\subsection{Taxonomy, Albedo, and Beaming}
Albedos and diameters derived from IR observations have proven useful for studies of the compositional diversity of asteroids throughout the Main Belt.  Previous measurements of the asteroids in the main belt suggested a uniform, smooth change in composition from more highly reflective, stony asteroids closer to the Sun to dark, carbon-rich objects that were barely altered by solar heating in its more distant regions.  However, data from the WISE mission and the Sloan Digital Sky Survey have shown that while there is an overall gradient, the true picture is much more jumbled, with a wide range of different compositions scattered throughout the belt.  Asteroids rich in minerals that require high temperatures to form have been found in the outer belt, while primitive objects are found close to its inner edge. These results suggest that the early migration of the giant planets may have turbulently scrambled the asteroids and show at the very least that the early solar system was more dynamic than previously believed \citep{DeMeo.2013a, Demeo.2014a}; see also DeMeo et al. (this volume).  

Comparisons between albedos derived from infrared data and taxonomic classifications derived from visible and near-infrared (VNIR) spectroscopy and spectrophotometry indicate a strong but not universal correlation between taxonomic types and albedos \citep{Mainzer.2011e, Mainzer.2011a, Thomas.2011a}.  Asteroids with neutral or blue spectra at VNIR wavelengths tend to have low visible albedos, and most red objects are associated with higher visible albedos (with the exception of classes such as D-types, which despite their red VNIR spectra have low albedos).  However, as described above, even though average \pv\ values may be nearly identical for some taxonomic types such as C-, P-, and D-types, their 3 $\mu$m albedos (\pIR) are distinctly different.  Therefore, \pIR\ can be used to distinguish between C/P types and D-types \citep{Grav.2012b} when VNIR spectra are unavailable.  Albedos have been used to identify candidate V-type asteroids throughout the Main Belt that have been subsequently observed spectroscopically to confirm their taxonomic type \citep{Hardersen.2014a}.  Moreover, objects with extremely low albedos (\pv$\sim$0.02) in the Main Belt have been identified as the possible progenitors of the Tagish Lake meteorite, one of the darkest and most primitive carbonaceous chondrites ever found \citep{Vernazza.2013a}. 

The beaming parameter $\eta$ has also been shown to correlate with thermal inertia and rotation rate \citep{Harris.1998a, Spencer.1989a, Delbo.2004a, Delbo.2007a}. However, caution must be used in the interpretation of beaming values, since $\eta$ also correlates with the phase angle at which an object is observed \citep{Delbo.2003a, Delbo.2007a, Wolters.2008a, Mainzer.2011d}.  The NEATM assumes a temperature distribution that drops to zero at the terminator.  For rotating bodies with anything other than low thermal inertias, the temperature distribution is likely to be non-zero past the terminator; when these bodies are observed at high phase angle, some portion of the night side flux is seen, resulting in changes to $\eta$ to conserve energy.  Moreover, increased surface roughness can increase the sunward beaming of radiation, and $\eta$ also depends on the spin axis orientation with respect to the Sun.  

With these caveats, objects with extremely high values of $\eta$ are good candidates for high thermal inertias or rapid rotation rates. \citet{Harris.2014a} found 18 NEOs with high $\eta$ values ($>$2) and albedos that may indicate a metallic composition consistent with the fragmented cores of differentiated bodies. Since high radar albedos can correspond to high metal content, follow-up radar observations are desirable to confirm their nature.  NEOs tend to have higher $\eta$ values than MBAs and Jovian Trojans; this may indicate higher thermal inertia, although they are often observed at higher phase angles than MBAs \citep{Delbo.2007a, Mainzer.2011d, Masiero.2011a, Grav.2011b}. \citet{Bauer.2013a} used WISE to perform thermal modeling on 52 Centaurs and scattered disk objects, finding that most $\eta$ values were low and similar to those of bare cometary nuclei, suggesting a common origin.  Observations of 85 trans-Neptunian objects by \citet{Lellouch.2013a} showed that high $\eta$ values were rare at low heliocentric distances, which they interpret as evidence of decreasing thermal inertia.  The advent of new mid-IR data from WISE, \emph{Spitzer}, AKARI, and \emph{Herschel} promises to shed further light on thermal inertia trends as thermophysical models are applied to many more objects (see Delb{\'o} et al., this volume).
 
\subsection{Activated Asteroids and Comets}
The boundary between asteroids and comets is now considerably more permeable than previously suspected.  Objects with low albedos and cometary orbits (often telltale signatures of comets) have stubbornly resisted attempts to find evidence of cometary activity such as comae or tails, and asteroids are sometimes revealed upon further inspection to have become active.  Thermal IR imaging surveys have been used to constrain cometary nucleus sizes during periods of inactivity when dust and gas do not obscure them, allowing for determinations of comet size frequency distributions \citep[e.g.][]{Bauer.2011a, Fernandez.2013a}.  Thermal IR imaging can also discover cometary activity around objects previously thought to be asteroidal.  For example, WISE observed extended emission around three asteroids during its prime mission, which were subsequently redesignated as comets (237/P LINEAR, 233/P La Sagra, and P/2009 WX51 Catalina), and \emph{Spitzer} found evidence for activity on (3552) Don Quixote \citep{Mommert.2014c}.  Since dust emits strongly at thermal IR wavelengths when comets reach the inner solar system, IR surveys can be used to discover new comets.  IRAS discovered five comets, and NEOWISE has discovered 24 to date.

Although a detailed discussion of comets is outside the scope of this book, thermal IR data can also constrain the sizes and quantities of the dust particles produced by comets \citep{Bauer.2012b}.  The slope of the particle size distributions in turn can be used to determine whether the dust was produced by a collision between two asteroids or driven off the surface by volatiles such as CO and CO$_{2}$, a key signature of cometary activity.  This technique is useful for understanding the nature of activity observed in some Main Belt asteroids \citep{Bauer.2012a, Stevenson.2012a}; see Jewitt et al. (this volume) for a detailed discussion of these objects.

\subsection{Mid-IR Spectroscopy of Asteroids}
A number of key spectral features that vary depending on a surface's composition, grain size, and porosity can be observed at mid-IR wavelengths.  Laboratory studies of meteorites indicate the presence of absorption and emission features caused by vibrational and lattice modes; for example, crystalline and amorphous silicates show broad absorption features near $\sim$10 $\mu$m.  

AKARI, \emph{Spitzer}, and \emph{Herschel} all carried spectrographs operating at mid-IR wavelengths during their fully cryogenic mission phases.  Spectra have been collected for $\sim$87 asteroids using the \emph{Spitzer} IRS and MIPS instruments \citep[e.g.][]{Cruikshank.2006a, Campins.2009a, Campins.2009b, Lim.2011a, Marchis.2012a}.  \emph{Spitzer} IRS spectra covering 5-38 $\mu$m revealed broad emission features near 10 and 20 $\mu$m on three Jovian Trojans, consistent with the presence of fine-grained silicates \citep{Emery.2006a, Cruikshank.2005a}.  Similar but smaller emission plateaus were observed between 9 - 12 $\mu$m on 8 Themis-family asteroids and (65) Cybele, indicating that these bodies may also be covered with small silicate grains \citep{Licandro.2011a, Licandro.2012a}; see also chapters by Delbo et al. and Emery et al. (this volume). Moreover, \citet{Vernazza.2012a} showed that mid-IR spectra meteorite samples diluted with IR-transparent KBr powder matched observed emissivity features in Main Belt asteroids, suggesting that surface porosity can also be constrained.  \citet{Marchis.2012a} used \emph{Spitzer} IRS spectroscopy of binary asteroids to constrain their bulk densities, thermal inertia, and surface grain properties.   The chapter by Reddy et al. (this volume) provides a more comprehensive discussion of mid-IR spectroscopy, including theory and observations.  

Mid-IR spectra have also proven useful for constraining the abundance of volatiles on small bodies; for example, many species such as CO and CO$_{2}$ have been identified on comets photometrically through excess flux above their thermal emission \citep[e.g.][]{Bauer.2011a, Reach.2013a}.  Recently, \emph{Herschel} heterodyne spectroscopy of the largest object in the Main Belt, (1) Ceres, has revealed the presence of water vapor as the dwarf planet approached perihelion \citep{Kueppers.2014a}.  The water vapor line was detected at three separate epochs at 557 GHz ($\lambda= 540 \micron$), further illustrating that the distinction between dark asteroids and comets is sometimes blurred \citep{Briani.2011a}.  See Rivkin et al. (this volume) for a discussion of the implications of this result.

\subsection{Earth Co-Orbitals}
Among the near-Earth objects discovered by NEOWISE during its post-cryogenic mission is the first known Earth Trojan, 2010 TK$_{7}$ and an object in a so-called ``horseshoe" orbit co-orbital with Earth, 2010 SO$_{16}$ \citep{Connors.2011a, Christou.2011a, Mainzer.2012c}.  2010 TK$_{7}$ was discovered by NEOWISE because its $\sim$395-year libration period caused it to move to the region near 90$^{\circ}$ solar elongation where the satellite continually observes.  Because Earth Trojans are thought to be dynamically constrained to remain more than $\sim24^{\circ}$ away from Earth in mean anomaly \citep{Tabachnik.2000a}, they spend most of their time in regions of the sky that are difficult or impossible for ground-based telescopes to observe.  Infrared measurements from NEOWISE have provided preliminary estimates of diameters and albedos \citep{Mainzer.2012c}.  2010 TK$_{7}$ is thought to be temporarily captured by Earth, with a dynamical stability timescale of $\sim$7000 years \citep{Connors.2011a, Marzari.2013a}.  By contrast, 2010 SO$_{16}$ librates across the L3 Earth-Sun Lagrange point in a horseshoe pattern that has the longest known stability of any Earth co-orbital, several hundred thousand years \citep{Christou.2011a}.  

Although approximately half a dozen horseshoe and quasi-satellite co-orbitals are known at present, 2010 TK$_{7}$ remains the sole Earth Trojan found to date.  These objects may represent a much larger population that remains undiscovered because they spend the majority of their time in the daytime sky on Earth.  Future surveys designed to survey at low solar elongations may be able to find more of these unusual objects.         

\section{Conclusions}
Space-based infrared studies of asteroids offer a valuable means of rapidly determining the physical and orbital properties of large numbers of objects.  Modern detector arrays have allowed recent space missions to reach background-limited sensitivities that are orders of magnitude improved compared to prior generations of IR telescopes. These capabilities have opened a new window onto the nature of our solar system's small bodies. Future generations of space-based IR telescopes using new large-format detectors \citep[e.g.][]{McMurtry.2013a} will further improve our understanding.

\section{ Acknowledgments.} This study is based on observations with AKARI, a JAXA project with the participation of ESA. This work is based in part on observations made with the \emph{Spitzer Space Telescope}, which is operated by the Jet Propulsion Laboratory, California Institute of Technology under a contract with NASA.  This publication makes use of data products from NEOWISE, which is a project of the Jet Propulsion Laboratory/California Institute of Technology, funded by the National Aeronautics and Space Administration.  This publication makes use of data products from the \emph{Wide-field Infrared Survey Explorer}, which is a joint project of the University of California, Los Angeles, and the Jet Propulsion Laboratory/California Institute of Technology, funded by the National Aeronautics and Space Administration.  \emph{Herschel} is an ESA space observatory with science instruments provided by European-led Principal Investigator consortia and with important participation from NASA.  We gratefully acknowledge the helpful inputs from referees M. Delbo and A. Harris of DLR. 

\bigskip


\end{document}